# Chunk List – Concurrent Data Structures

## Daniel Szelogowski

**Abstract**— Chunking data is obviously no new concept; however, I had never found any data structures that used chunking as the basis of their implementation. I figured that by using chunking alongside concurrency, I could create an extremely fast run-time regarding particular methods such as searching and/or sorting. By using chunking and concurrency to my advantage, I came up with the chunk list — a dynamic list-based data structure that would separate large amounts of data into specifically sized chunks, each of which should be able to be searched at the exact same time by searching each chunk on a separate thread. As a result of implementing this concept into its own class, I was able to create something that almost consistently gives around 20x-300x faster results than a regular ArrayList. However, should speed be a particular issue even after implementation, users can modify the size of the chunks and benchmark the speed of using smaller or larger chunks, depending on the amount of data being stored.

**Index Terms**—Distributed data structures, parallelism and concurrency, parallel databases, conversion from sequential to parallel forms

◆

## 1 INTRODUCTION

### 1.1 What is a Chunk List?

A chunk list is an array-based list of elements in which data is stored in inner lists of a certain capacity, allowing for easily modifiable and faster runtimes based on the number of elements being stored. A simple way to conceptualize a chunk list would be an ArrayList (dynamic array) of ArrayLists. The main list would contain the "chunks", or ArrayLists that are not allowed to be filled past a specific capacity. Any time a "chunk" has reached capacity, a new ArrayList is added, and items are added to that chunk from thereon.

By doing this process and splitting our list into chunks, we can use parallel processing to our advantage. Using concurrency, we can run each chunk on a separate thread when doing tasks such as searching or removing.

This can be expressed visually as a table: as an example, a chunk list containing the numbers $1 - 50$ where the chunk size is set to 10 elements. (See **Fig. 1**)

| [0] | | | [1] | | | [2] | | | [3] | | | [4] | | |
|----|----|----|----|----|----|----|----|----|----|----|----|----|----|----|
| 1  | 2  | 3  | 11 | 12 | 13 | 21 | 22 | 23 | 31 | 32 | 33 | 41 | 42 | 43 |
| 4  | 5  | 6  | 14 | 15 | 16 | 24 | 25 | 26 | 34 | 35 | 36 | 44 | 45 | 46 |
| 7  | 8  | 9  | 17 | 18 | 19 | 27 | 28 | 29 | 37 | 38 | 39 | 47 | 48 | 49 |
| 10 | | | 20 | | | 30 | | | 40 | | | 50 | | |

**Fig. 1.** Visual representation of a Chunk List.

### 1.1.1 Defining a Chunk

Relatively speaking, "chunking" as a concept can be defined similarly to its psychological definition: Chunking is a term referring to the process of taking individual pieces of information (chunks) and grouping them into larger units [1]. Essentially, we are dividing up data into multi-

- *Daniel Szelogowski is with the Department of Computer Science, University of Wisconsin at Whitewater, WI 53190. E-mail: szelogowdj19@uww.edu*

ple partitions to manipulate each one concurrently. One chunk may contain a large or small portion of our dataset, depending on how we want the elements to be partitioned and the amount of data being stored collectively across the entire structure.

### 1.1.2 Efficiency of Chunking

Chunking data for purposes of efficiency is a highly common practice. To compare, network optimization utilizes the same idea in the form of packets: On the Internet, the network breaks an [data/messages] into parts of a certain size in bytes. These are the packets [2].

### 1.2 Where is a Chunk List Used?

The basis of the data structure makes it useful for storing very large and very small amounts of elements. Benefits shine especially when a list is unsorted:

- Fast searching.
- Fast removal.
- Fast insertion.
- Fast iteration.

In any scenario, a chunk list can be used in place of an ArrayList especially, as well as something such as a Binary Search Tree, as searching may be faster based on processing power.

### 1.3 Benefits

Implementation is easy and short, and sorting is quick even with large amounts of chunks and optimization. With the ease of adjustability of chunk size, the capacity can be modified to allow for faster and more efficient speeds.

### 1.3.1 Area-Specific Usage

The most likely real-life scenario in which a chunk link would be preferred are for video games and optimization – particularly in the sense that many video games today, especially sandbox-style games such as Minecraft use a process called 'chunking' for map data – combining and decompiling maps into 'chunks' in order to load only the parts of the map within a radius of the player for the pur-pose of increased performance by reducing the amount of entities loaded within a visible area [5].

## 2 IMPLEMENTATION DETAILS

### 2.1 Construction

The basis of the chunk list is the inner list. This is best implemented using some sort of dynamic list, such as ArrayList (or List in C#).

This inner list will start out with a single list on the inside. Constructor must include an integer, the chunk size. Otherwise, revert to a default size.

New lists (chunks) will only be added to the main list when the chunk at the end has reached capacity. Likely, the best implementation for a constructor would be to set the chunk size to the square root (as an integer) of the amount of data being stored, as in testing this has yielded the fastest performance. [4] This is the most sensible size to use particularly for the resulting computational time: our aim is to obtain Big-O (log n) of some sort, which is especially the case as a result of pre-dividing up our data into a logarithmic section.

A chunk list may be implemented with generics (or templates) so long as the generic type is comparable.

---
**Algorithm 1** Class Definition

```
class {
    List< List< T> > myList ;
    int chunkSize ;
    const int DEFAULT_SIZE ← 1000 ;
} ChunkList< T> where T implements IComparable
```
---

### 2.2 Multithreading Methods

Multithreading is an especially important part of chunk list implementation, as the basis of the list's speed is primarily the result of concurrency. For most methods in a chunk list, a new thread can be created for each chunk to be iterated through.

A good example of this lies within C#'s Parallel.ForEach method, which will be referred to for this type of operation.

Thread synchronization is not required when iterating, however keeping track of the thread state is important in some instances.

### 2.3 Index-Based Methods

Accessing or modifying an element at a specified index (such as get, set, or removeAt methods) is somewhat more complex than in a regular list.

To get the chunk where the position would be located, divide the index by the chunk size and cast it to an integer: $chunk = int(index\ /\ chunkSize)$

This method will work regardless of the current capacity of the chunk, given that we are accessing the chunk relative to the span of the list. The same applies for the index within the chunk that we need to access. Should this seem to be an issue, one could simply step down by one index to avoid a null index, or simply throw an error that the index contains no data (yet).

To get the position in the chunk where the index would be, use modulo on the index by the chunk size: $chunkPosition = index\ \%\ chunkSize$

We can then access the data via $list[chunk][chunkPosition]$ (Where list is the main list inside the class).

---
**Algorithm 2** Index-Accessing

```
private function CONVERTINDEXTOCHUNK(int index):
    return index / chunkSize ;
end function

private function CONVERTINDEXTOCHUNKPOS(int index):
    return index MOD chunkSize ;
end function
```
---

| Index Example | | | | |
|---|---|---|---|---|
| **Chunk 0** | | | | |
| [0] | [1] | [2] | [3] | [4] |
| 0 | 1 | 2 | 3 | 4 |
| **Chunk 1** | | | | |
| [0] | [1] | [2] | [3] | [4] |
| 5 | 6 | 7 | *8 | 9 |
| **Chunk 2** | | | | |
| [0] | [1] | [2] | [3] | [4] |
| 10 | | | | |
| **Accessing the chunk:** int(8 / 5) = 1 | | | | |
| **Accessing the chunk position:** 8 % 5 = 3 | | | | |

**Fig. 2.** Demonstration of accessing an element at index 8 in a chunk list containing numbers 0 – 10 with chunk size 5.

### 2.3.1 Index Issues

One issue with using indices in a chunk list, however, is the problem where items flow left (step down by index until falling in place with the rest of the list, as in the style of a linked list) within the chunk but do not migrate left from one to another (pulling items from the next chunk to fill the previous) if a chunk has an open slot. To implement so may hinder performance during removal.

However, a very simple solution would be to use recursion, such as within a try-catch statement using the index + 1.

*The following example demonstrates a solution to the problem by counting up the index until an open position is found, or throwing an error if the index is unreachable or beyond the span of the list:*

```
Algorithm 3 Potential Access Solution

public function GET(int index):
    if index ≥ SIZE(): # Size is the number of all items in list (total length)
        THROW new ARGUMENTOUTOFRANGEEXCEPTION() ;
    end
    try:
        return  myList[CONVERTINDEXTOCHUNK(index)][CONVERTINDEXTOCHUNKPOS(index)]  ;
    catch ARGUMENTOUTOFRANGEEXCEPTION():
        return GET(index + 1) ;
    end
end function
```

### 2.3.2 Chunk Resizing

Should our data set grow marginally larger, we may need to resize our list. To do so however, means we will need to rebalance our list, which is especially important if the chunk size we are changing to is smaller than the current one.

We can make a temporary list containing all our old items, change the chunk size, clear our old list, and then reflow our data back in.

While somewhat costly performance-wise, this is an operation that should not be necessary to occur often.

If the chunk size we want to adjust to is larger than the current one, however, we can simply leave the list as is and allow the elements to re-fill the chunks that are not yet at capacity.

```
Algorithm 4 Chunk Resizing

public function SETCHUNKSIZE(int index):
    if newChunkSize > chunkSize:
        chunkSize ← newChunkSize ;
    else
        items ← GETLIST() ; # Get ChunkList as ArrayList
        chunkSize ← newChunkSize ;
        CLEAR() ; # Clear the entire list
        foreach item in items:
            ADD(item) ;
        end
    end
end function

public function GETLIST():
    items ← new List< T > () ;
    foreach currentItem in myList:
        foreach currentItem in currentList:
            items.add(currentItem) ;
        end
    end
end function
```

### 2.4 Element Operations

We should include the standard operations equivalent to an array list. Manipulating the data will work very similarly but will focus more on the data chunks and the usage of concurrency for each method.

### 2.4.1 Adding Elements

Adding elements to a chunk list is simple; however, it does require that we check if each chunk is at capacity. Getting the size from the chunk should be Big-O (1), so this should not increase runtime marginally whatsoever.

An element will naturally fall into the first open spot, or the first chunk that is not at capacity. (See **Fig. 3**)

If all chunks are at capacity, however, we need to add a new chunk to our list, then add the item it. The resulting time would be processed in the time of Big-O (log C).

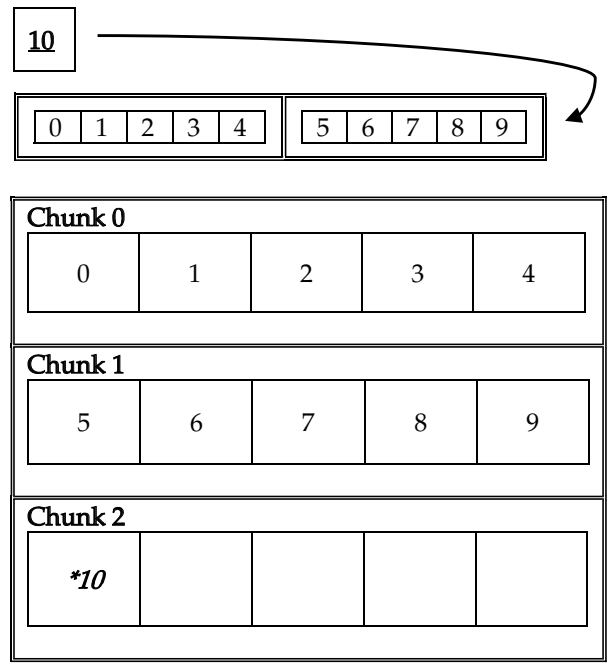

**Fig. 3.** Demonstration of adding an element to the end of a Chunk List with two currently full chunks, creating a third.

```
Algorithm 5 Add Items to List

public function ADD(T t):
    foreach currentList in myList:
        if currentList.COUNT = chunkSize:
            currentList.ADD(t) ;
            return
    end
    myList.ADD(new  List< T > ()) ;
    myList[myList.Count −1].ADD(t) ;
end function
```

### 2.4.2 Removing Elements

Removing elements is one of the fastest computational operations in a chunk list. This is where we can start using multithreading to our advantage.

To remove an element, we can use a parallel for loop to concurrently check each chunk for the item.

We can use a binary search to get the index that we're looking for.

This is also where we need to be able to have access to the thread's state when we are looping through each chunk. If we only want to remove the first found instance of an element, we need to immediately break out of the parallel for loop.

To remove all instances of an element within the list, we can still use a parallel for loop, and just call a removeAll method on each chunk. Given a more powerful computer, these events should both be fairly fast: our average time should be Big-O ((log C * log N) / P), and at worst we are only losing our divisor of processor cores at Big-O (log C * log N), based on the division of threads for each removal operation.

To clear the entire list, we can simply call clear on the main list (containing the chunks).

---

**Algorithm 6** Remove Items from List

---

*public* **function** REMOVE(T *t*):
  Parallel.FOREACH(*myList* , (*currentList* , *state*) →
    *index* ← *currentList*.BINARYSEARCH(*t*) ;
    **if** *index* ≥ 0:
      *currentList*.REMOVEAT(*index*) ;
      *state*.BREAK() ;
    **end**
  )
**end function**

*public* **function** REMOVEALL(T *t*):
  Parallel.FOREACH(*myList* , (*currentList*) →
    **for** *i* ←0 **to** *currentList*.COUNT −1:
      **if** *currentList*[*i*] = *t*:
        *currentList*.REMOVEAT(*i*) ;
        *i*−− ;
      **end**
      *i*++ ;
    **end**
  )
**end function**

---

### 2.4.3 Searching

Searching for an element is also where chunk lists shine. Once again, we can use concurrency to get the shortest possible runtime, as now we can use a parallel for loop not only on the list itself, but on each chunk.

Essentially, we can check most items in the list at the exact same time, meaning our runtime will be marginally smaller than using a linear search at worst case, and in the best case, a binary search. This is the result of our parallel search form: by opening each chunk on a separate thread, our goal is for one of our chunks to be successfully binary searched, even completely through without having the same linear performance for the rest of the list. Resulting is our more-likely Big-O ((log C * log N) / P), but should we have to search the entirety of every chunk for the full list, we may fall into the computational span of linear time Big-O (C * N). Of course, this should also be lessened by the number of threads opened, preventing a completely consecutive search time.

---

**Algorithm 7** Search

---

*public* **function** *contains*(T *t*):
  bool *found* ← **false** ;
  Parallel.FOREACH(*myList* , (*currentList* , *state*)→
    Parallel.FOREACH(*currentList* , (*currentItem*) →
      **if** *currentItem* = *t*:
        *found* ←**true** ;
        *state*.BREAK() ;
      **end**
    )
  )
  **return** *found* ;
**end function**

---

### 2.4.4 Sorting

Sorting our list is a fairly complex operation; to properly sort our list, we do have to make a temporary list containing all elements of our chunk list. To do otherwise would only sort the chunks, which is not ideal as we do not know which order they will be inserted in.

Using our temporary list, we can clear our main list and simply reflow all of our items back in after sorting it.

---

**Algorithm 8** Sorting

---

*public* **function** SORT():
  *items* ← GETLIST() ; # Get *ChunkList* as ArrayList
  *items*.SORT() ; # See below
  CLEAR() ;
  **foreach** *item* **in** *items*:
    ADD(*item*) ;
  **end**
**end function**

*public* **function** ArrayList.SORT():
  **if** *partition size is fewer than 16 elements*:
    INSERTIONSORT() ;
  **else if** *number of partitions exceeds 2 * log N , where N is the range of the input array*:
    HEAPSORT() ;
  **else**:
    QUICKSORT() ;
  **end**
**end**

---

## 3  COMPLEXITY ANALYSIS

### 3.1 Element-Based Methods

We can find the computational complexities by comparing those of a standard abstract list and dividing up the data based on the equivalent methods and the chunks of data as individual lists, acting as the size of each sub-list containing its own number of elements. In instances where we see C * N for example, this would represent either the entire list, or log C * log N representing a divisional portion of the data for the sake of computation complexity.

TABLE 1
COMPLEXITY OF BASIC METHODS

| Operation | Average Case | Worst Case |
|---|---|---|
| Add | O(1) | O(log C) |
| Remove | O((log C * log N) / P) | O(log C * log N) |
| RemoveAll | O((log C * N) / P) | O(log C * N) |
| RemoveAt | O(1) | O(C * N - I) |
| Set | O(1) | O(C * N - I) |
| Get | O(1) | O(C * N - I) |

*Complexities are listed with the following variables:*

-   *C being the number of chunks currently in the list.*
-   *N being the number of elements per chunk.*
-   *P being the number of processors.*
-   *I being the index input for the operation.*

### 3.2 List-Based Methods

These methods have been computed based primarily off instant computations or the computation complexity of the base method being used (e.g., the Sort method being derived from the built-in Sort method as mentioned in **Section 2.4.4**).

TABLE 2
COMPLEXITY OF ADDITIONAL LIST METHODS

| Operation | Average Case | Worst Case |
|---|---|---|
| GetList | O(C ^ 2 * N) | N/A |
| Contains (Search) | O((log C * log N) / P) | O(C * N) |
| Size (Count) | O(C) | N/A |
| SetChunkSize | O(1) | O(C ^ 2 * N) |
| Sort | O(C * N * log N) | O(C * N ^ 2) |

## 4 USAGE AND EXAMPLES

### 4.1 Modern Usage

The biggest potential usage for the chunk list would definitely be for video games – any time a large number of objects or map data would need to be contained and searched through, the data structure would provide the most efficient mean to load portions of data as well as find objects within the chunks.

### 4.1.1 Unit Test and Benchmarks

A working unit test can be found on the GitHub repository [4] comparing results of data computations using a chunk list, a chunk list with the chunk size set to the square root of the data quantity, and a standard array list.

### 4.1.2 Competition

Many data structures could potentially either replace or be replaced by the chunk list – in smaller amounts of data, an array list could of course be used, or even a binary tree if searching through the data is the key important part. The biggest selling point is of course the concurrency: any time speed is the biggest factor in manipulating data, the chunk list is a likely competitor compared to a graph or binary search tree. As such, this structure would work especially well as the backend of a database management system, in place of B-trees, for example.

### 4.2 Related Work

As mentioned previously, Minecraft's chunking system was a drawn upon concept in development of the data structure. The idea of chunking data into smaller sets in order to access the most important parts and use different threads and processor cores to manipulate data concurrently provided a baseline to the core design of the structure. Cryptocurrencies and other mathematical challenges such as Project Euler also were key to finding an importance in speed in data manipulation, as the usage in problem solving and video game optimization were two key ideas meant to be solved through the implementation of the data structure.

# APPENDIX
*Method Headers & Complexities*

| OPERATION | AVERAGE CASE | WORST CASE |
|---|---|---|
| **Constructor** *()* : *this(DEFAULT_SIZE)* | *N/A* | *N/A* |
| **Constructor** *(int chunkSize)* | *N/A* | *N/A* |
| void **Add** *(T t)* | O(C) | *N/A* |
| void **Remove** *(T t)* | O((log C * log N) / P) | O(log C * log N) |
| void **RemoveAll** *(T t)* | O((log C * N) / P) | O(log C * N) |
| void **RemoveAt** *(int index)* | O(1) | O(C * N - I) |
| int **ConvertIndexToChunk** *(int index)* | *N/A* | *N/A* |
| int **ConvertIndexToChunkPos** *(int index)* | *N/A* | *N/A* |
| void **Set** *(int index, T t)* | O(1) | O(C * N - I) |
| T **Get** *(int index)* | O(1) | O(C * N - I) |
| List<T> **GetList** *()* | O(C ^ 2 * N) | N/A |
| bool **Contains** (Search) *(T t)* | O((log C * log N) / P) | O(C * N) |
| int **Size** (Count) *()* | O(C) | *N/A* |
| void **SetChunkSize** *(int size)* | O(1) | O(C ^ 2 * N) |
| void **Sort** *()* | O(C * N * log N) | O(C * N ^ 2) |